\documentclass[prd,preprint,groupedaddress,nofootinbib,showpacs,preprintnumbers,11pt]{revtex4}

\pdfoutput=1

\usepackage{epsfig}
\usepackage{hyperref}
\usepackage{amssymb}
\usepackage{amsbsy}
\usepackage{amsmath}
\usepackage{url}

\newcommand{\rf}[1]{(\ref{#1})}
\newcommand{\beq}{\begin{equation}}
\newcommand{\eeq}{\end{equation}}
\newcommand{\be}{\begin{equation}}
\newcommand{\ee}{\end{equation}}
\newcommand{\bea}{\begin{eqnarray}}
\newcommand{\eea}{\end{eqnarray}}
\newcommand{\eq}[1]{Eq.~(\ref{#1})}
\newcommand{\non}{\nonumber \\*}
\newcommand{\ie}{{i.e.}\ }
\newcommand{\bv}{\begin{verbatim}}
\newcommand{\ev}{\end{verbatim}}

\newcommand{\vp}{\varphi}

\newcommand{\e}{\,\mbox{e}}
\renewcommand{\d}{{\rm d}}
\renewcommand{\i}{{\rm i}}
\newcommand{\blambda}{\bar\lambda}
\newcommand{\brho}{\bar\rho}

\newcommand{\bz}{{\bar z}}
\newcommand{\p}{\partial}
\newcommand{\bp}{\bar\partial}
\newcommand{\q}{\mbox{}}

\newcommand{\eps}{\varepsilon}
\newcommand{\om}{\omega}

\newcommand{\LA}{\left\langle}
\newcommand{\RA}{\right\rangle}

\def\fun#1#2{\lower3.6pt\vbox{\baselineskip0pt\lineskip.9pt
\ialign{$\mathsurround=0pt#1\hfil##\hfil$\crcr#2\crcr\sim\crcr}}}

\begin{document}


\title{Opus on conformal symmetry of the Nambu-Goto\\[1mm] versus Polyakov strings}

\author{Yuri Makeenko}
\address{NRC ``Kurchatov Institute''\/-- ITEP, Moscow\\
\vspace*{1mm}
{\email: makeenko@itep.ru} }

\begin{abstract}
I investigate the Nambu-Goto and Polyakov strings, accounting for higher-derivative
terms in the emergent action for the metric tensor which are classically negligible for smooth
metrics but revive quantumly. Using the conformal field theory technique by KPZ-DDK,
I~compute in the one-loop approximation the conformal dimension and the central charge
which differs in the two cases,
telling the Nambu-Goto and Polyakov strings apart.
I confirm the results by explicit quantum field theory computations
of the propagator and the energy-momentum tensor
at one loop, using the Pauli-Villars regularization.

\end{abstract}


\maketitle

\begin{flushright}
\parbox[t]{5.8cm}
{Salieri:  \mbox{}\\
{\small ``I checked the harmony with algebra.\\[-1mm]
Then finally proficient in the science, \\[-1mm]
I~risked the rare delights of creativity.''}
\\[.2cm]
{\sc A. Pushkin}, {\it Mozart and Salieri}}   
\end{flushright} 
\vspace*{1.cm}

 \section{Introduction}

It is commonly believed that the Nambu-Goto and Polyakov strings 
are equivalent except the dimension of the target space 
parametrized by $X^\mu$ is shifted
$d\to d-1$ because the metric $g_{ab}$ is independent for 
the Polyakov string formulation~\cite{Pol81}.
The latter is described by the Brink-Howe-Di Vecchia-Polyakov action
\be
S=\frac1{4\pi \alpha'}\int \sqrt{g} g^{ab} \p_a X \cdot \p_b X
\label{SPol}
\ee
which is quadratic in $X^\mu$ that makes it easy to integrate it out
in the path integral. Alternatively, the Nambu-Goto action is the area
of the string worldsheet which is highly nonlinear in $X^\mu$ but
can be made quadratic, introducing the Lagrange multiplier 
$\lambda^{ab}$ and an independent metric tensor $g_{ab}$,
\be
S_{\rm NG}=\frac1{2\pi \alpha'}\int \sqrt{\det {(\p_a X\cdot \p_b X)}}=
\frac1{2\pi \alpha'}\int \Big (\sqrt{g} +\frac 12\lambda^{ab} (\p_a X\cdot \p_b X-g_{ab}) \Big).
\label{SNG}
\ee

The equivalence of the two string formulations is evident at the classical level.
It was also demonstrated~\cite{FTs82} at one loop about the classical string ground state.
The general argument in favor of the equivalence is based~\cite{Pol87} on the fact that
$\lambda^{ab}$ propagates only to the distances of the order of the ultraviolet cutoff and
therefore should not affect the theory at large distances.
However, a more careful analysis of the Nambu-Goto string shows~\cite{AM16} that the true
ground state of the strings with simplest topology of disk, cylinder 
or torus occurs at a certain value
\be
\lambda^{ab}=\blambda \sqrt{g} g^{ab},\qquad  1/2<\blambda<\blambda_{\rm cl}=1.
\label{mf}
\ee
It is stable at  the mean-field value $\blambda$
for $2<d<26$ in contrast to the classical value $\blambda_{\rm cl}=1$ which is stable only for $d<2$ where
it results in the Polyakov string. Analogously, the mean-field value of the metric tensor 
\be
g_{ab}=\brho \delta_{ab},\qquad \brho>\brho_{\rm cl}
\label{mfg}
\ee 
in the conformal gauge, where $\brho$ is constant for the worldsheet parametrization and 
$\brho_{\rm cl}$ stands for the classical induced metric. Nevertheless,
the discrepancy of $\brho$ from $\brho_{\rm cl}$ does not  mean that  the 
Nambu-Goto and Polyakov string are physically different because of the so-called background
independence which stays in the given case that $\brho$ is not observable thanks to
the Weyl invariance.

The goal of this Paper is to investigate the Nambu-Goto versus Polyakov strings at one loop about
the ground states, using the methods of conformal field theory.
To be precise I am going to verify for the Nambu-Goto string the celebrated formula
\be
\gamma_{\rm str}=(1-h)\left[\frac{d-25-\sqrt{(25-d)(1-d)}}{12}\right]+2
\label{ggg}
\ee
by Knizhnik-Polyakov-Zamolodchikov~\cite{KPZ}  and
David-Distler-Kawai~\cite{Dav88} (abbreviated as KPZ-DDK)
for the string susceptibility index $\gamma_{\rm str}$ --
also known as
the gravity anomalous dimension -- of a surface of genus $h$. It was derived
for the Polyakov string and suffers the so-called $d=1$ barrier above which \rf{ggg} is not real
and thus unacceptable.

The computation will be performed using the emergent (or induced) action for $g_{ab}$
\be
{\cal S}=\frac 1{16 \pi b_0^2} \int \sqrt{g} \left[ - R \frac 1\Delta R +2m_0^2
+a^2  R \left(R+G g^{ab}\,
\partial_a \frac 1\Delta R\;\partial_b \frac 1\Delta R \right)
\right], 
\label{inva}
\ee
with $b^2_0=6/({26-d})$,
which emerges after path integration over $X^\mu$, ghosts and $\lambda^{ab}$.
In \eq{inva}
$R$ is the scalar curvature associated with $g_{ab}$, $\Delta$ is the two-dimensional Laplacian
and $a$ is an ultraviolet target-space cutoff. 
For the Polyakov string the action \rf{inva} with $G=0$ 
is derivable from 
the Seeley expansion of the heat kernel~\cite{deWitt}.   
The leading-order term is associated with the
conformal anomaly~\cite{Pol81} while the $R^2$ term comes from the next order in $a^2$.
It is familiar from the studies~\cite{KN93,Ich95,KSW} of the $R^2$ two-dimensional gravity,
where the first term on the right-hand side of the action \rf{inva} was missing.
The emergence of the second higher-derivative term with $G\neq 0$ on the right-hand side of \eq{inva}
for the Nambu-Goto string was demonstrated by the present author~\cite{Mak21}.
The actual value of $G$ was not computed but it was shown to be nonvanishing
for the Nambu-Goto string.

The additional terms  in the action~\rf{inva} are suppressed for
smooth metrics as powers of $a^2 R$. However, typical metrics which are essential in the
path integral over the metrics $g_{ab}$ are not smooth and have $R\sim a^{-2}$, 
so the higher-derivative terms revive~\cite{Mak21} after doing uncertainties like $a^2 \times a^{-2}$. 
The first term in the action~\rf{inva} is thus
only an effective action governing smooth classical configuration, while 
the higher-order terms result in nontrivial interactions.
There are also other higher-derivative terms of the order $a^4$ and higher but it was argued~\cite{Mak21}
they probably do not change the results owing to the universality.

In the present Paper I perform the standard computation of the central charge and
the conformal dimension of $\sqrt{g}$ at one loop,  
taking into account the higher-derivative terms in the action~\rf{inva}.
The loop expansion goes in $b_0^2$ which is treated as the parameter of a semiclassical
expansion. The obtained result for the string susceptibility index
in the one-loop approximation reads
\be
\gamma_{\rm str}=
(1-h)\left( \frac {1}{b_0^2}  -\frac76-G +{\cal O}(b_0^2)\right) +2 .
\label{gstrfin}
\ee
The most interesting in this result is its dependence on $G$. If $G=0$, when the action
\rf{inva} is simply the action of the Polyakov string with the curvature-squared added, 
then \eq{gstrfin} reproduces 
for $b_0^2=2/(26-d)$ the expansion of \rf{ggg}. However, 
 for the Nambu-Goto string where $G\neq 0$ we see in \rf{gstrfin} the deviation from
the expansion of  \rf{ggg} showing up already at one loop.

The Paper is organized as follows. In Sect.~\ref{s:em} I derive the energy-momentum tensor
associated with the action \rf{inva}. It is conserved and traceless in spite of the presence
of the dimensional parameter $\eps$ thanks to diffeomorphism invariance.
In Sect.~\ref{s:III} the formulas are rewritten in the conformal coordinates.
Sect.~\ref{s:G0} is devoted to DDK for the Polyakov string with curvature squared,
\ie $G=0$.
The results at one loop are the same as for the usual Polyakov string from textbooks.
In Sect.~\ref{s:V} I extend DDK to the Nambu-Goto string and show that the central 
charge gets additionally $6G$ at one loop. In Sect.~\ref{s:sal} I check
the results by explicit one-loop computations, using the Paul-Villars regularization.
I show that only the renormalization of $\mu_0^2$ is infinite, while all other
divergences, including logarithmic, are mutually canceled in physical quantities, 
thus preserving conformal invariance. 
The remaining finite parts precisely confirm the additional $6G$ in the central charge.
In Sect.~\ref{s:conclu} I briefly discuss the results.
Some useful formulas with the regularized propagator are collected in Appendix~A.


\section{The energy-momentum tensor\label{s:em}}

We begin with fixing the conformal gauge 
\be
g_{ab}= \brho  \e^\vp  \hat g _{ab},
\label{confog}
\ee
where  $\hat g_{ab}$ is the fiducial metric tensor and $\vp$ is the dynamical variable
often called the Liouville field. In the gauge  \rf{confog} the action \rf{inva} becomes
\bea
 S &=& \frac1{16 \pi b_0^2}\int \sqrt{\hat g}\left[-\left(\q\hat R- {\hat \Delta} \vp \right)
\frac{1} {\hat \Delta}\left(\q\hat R- {\hat \Delta}\vp \right) +2\mu_0^2
+\eps\left(\q\hat R -\hat \Delta \vp\right)^2 \right.\non
 && \left.+G \eps  \e^{-\vp} \left(\q\hat R -\hat \Delta \vp \right)
\hat g^{ab} \partial_a \left(\frac{1\q}{\hat \Delta} \hat R- \vp \right)
\partial_b \left(\frac{1\q}{\hat \Delta} \hat R- \vp \right)\right],
\label{A13}
\eea
where $\eps=a^2/\brho$ plays the role of the worldsheet cutoff, 
$\mu_0^2=m_0^2\brho$ and we used the relation
\be
\sqrt g R= \sqrt{\hat g} \left( \q\hat R - \hat \Delta \vp \right)
\label{Rshift}
\ee
between the scalar curvatures $R$ and $\hat R$ for 
the metrics $g_{ab}$ and $\hat g_{ab}$, respectively.
The two-dimensional Laplacian reads
\be
\Delta\equiv\frac 1 {\sqrt{ g}}  \partial_a \sqrt{ g}  g^{ab} \partial _b
=\e^{-\vp} \frac 1 {\sqrt{\hat g}} \partial_a \sqrt{\hat g} \hat g^{ab} \partial _b \equiv
\e^{-\vp} \hat \Delta
\label{defDel}
\ee
in  the gauge \rf{confog}.

The energy-momentum tensor $T_{ab}$ can be derived in the standard way by varying 
\rf{A13} with respect to $\hat g^{ab}$. 
If $\vp$ was just a scalar field ``minimally'' coupled to gravity described by 
the  metric tensor $\hat g_{ab}$,  the action would be simply given by \rf{A13},
where we would set $\hat R=0$.
We would then obtain 
\bea
-4 b_0^2 T_{ab}^{({\rm min})}&=&\p_a \vp \p_b \vp -\frac 12 g_{ab} \p^c \vp \p_c \vp
 -\mu_0^2 g_{ab}-\eps \p_a \vp \p_b \Delta \vp -
\eps \p_a \Delta \vp  \p_b \vp \non && +\eps g_{ab} \partial^c \vp\p_c \Delta \vp+\frac \eps2 g_{ab} (\Delta \vp)^2 
-G \eps \p_a \vp\p_b\vp \Delta \vp 
 + G\frac\eps2\p_a\vp \p_b(\p^c \vp \p_c \vp) \non &&
+ G\frac\eps2 \p_a (\p^c \vp \p_c \vp)\p_b\vp -G\frac\eps2
g_{ab} \p^c\vp \p_c (\p^d \vp \p_d \vp) ,
\label{Tmin}
 \eea
 where the indices are raised by $g^{ab}$.
It is conserved owing to the classical equation of motion 
\bea
&&-\Delta \vp+\mu_0^2 + \eps \Delta^2 \vp -\frac{\eps}2 (\Delta \vp)^2  
+\frac12 G\eps \p^a \vp \p_a \vp \Delta \vp-\frac12 G\eps \p^a\p_a(\p^b\vp\p_b\vp)\non
&&\hspace*{.85cm}+
 G\eps \p_a  (\p^a \vp \Delta \vp) =0
\label{cemG}
\eea
but is not traceless:
\be
\partial^a T_{ab}^{(\rm min)}=0,\qquad  T_{a}^{a\,(\rm min)}\equiv 
 g^{ab}T_{ab}^{(\rm min)}=\frac1{2b_0^2}\left[\mu_0^2
-\frac\eps2 \Delta \vp ( \Delta \vp  - G \p^a \vp \p_a \vp )\right].
\ee

Keeping the conservation, the trace can be nullified by adding a nonminimal coupling to gravity
as written in the action \rf{A13}.
This is a general property of theories with diffeomorphism invariance. 
The corresponding energy-momentum tensor can be derived by 
varying with respect to  $\hat g^{ab}$, using the standard formulas
\be
\frac {\delta \sqrt{\hat g}}{\delta \hat g^{ab}}=
-\frac 12 \sqrt{\hat g} \hat g_{ab} ,\qquad
\frac {\delta \sqrt{\hat g}\hat R}{\delta \hat g^{ab}}=
-2 \left(\partial_a\partial_b-\hat g_{ab} \hat g^{cd}\partial_c \partial_d \right),
\ee
where after the variation we set $\hat g_{ab}$ to be constant. We obtain%
\footnote{Our approach is like in Ref.~\cite{DJ95} for the Liouville action.}
\bea
-4 b_0^2 T_{ab} &=&-4 b_0^2 T_{ab}^{({\rm min})}
-2 \q(\p_a\p_b-g_{ab}\p^c\p_c )(\vp -\eps \Delta \vp+G\frac\eps2 g^{ab}\p_a \vp \p_b \vp  ) 
\non &&
 + 2 G \q\eps (\p_a \p_b-g_{ab} \p^c\p_c) 
 \frac 1{\Delta} \partial_d(\p^d \vp \Delta\vp)
\eea
or more explicitly
\bea
-4 b_0^2 T_{ab}
&=&\p_a \vp \p_b \vp -\frac 12 g_{ab} \p^c \vp \p_c \vp -\mu_0^2 g_{ab}
-\eps \p_a \vp \p_b \Delta \vp -
\eps \p_a \Delta \vp  \p_b \vp 
\non&& +\eps g_{ab} \partial^c \vp\p_c \Delta \vp+\frac \eps2 g_{ab} (\Delta \vp)^2
-G \eps \p_a \vp\p_b\vp \Delta \vp
 + G\frac \eps2\p_a\vp \p_b(g^{cd}\p_c \vp \p_d \vp)  \non
 && 
+  G\frac\eps2\p_b\vp \p_a (g^{cd}\p_c \vp \p_d \vp) - G\frac\eps2
g_{ab} \p^c\vp \p_c (g^{de}\p_d \vp \p_e \vp) 
\non
 &&
-2\q(\p_a\p_b-g_{ab}\p^c\p_c )(\vp -\eps \Delta \vp+G\frac\eps2 g^{cd}\p_c \vp \p_d \vp  ) 
\non &&
 + 2 G \q\eps (\p_a \p_b-g_{ab} \p^c\p_c) 
 \frac 1{\Delta} \partial_d(g^{de}\p_e \vp \Delta\vp). 
 \label{Tab}
\eea

The energy-momentum tensor \rf{Tab}
is conserved and traceless thanks to the classical equation of motion \rf{cemG}.
For $G=0$ it generalizes the symmetric energy-momentum tensor
of a higher-derivative scalar field \cite{scalar}
to the theory with diffeomorphism invariance.

Notice the nonlocality of the last term in \rf{Tab} which is inherited from a
nonlocality of the invariant action \rf{inva}. As we shall see below, the presence of this nonlocal term will be crucial to provide the difference between the Nambu-Goto and Polyakov strings 
at one loop.

It worth also mentioning that the trace 
$ 
T_a^a=0
$ 
classically despite the presence of
the dimensionful worldsheet cutoff $\eps\sim a^2/\brho$ thanks to the two last terms in \rf{Tab} 
which come from diffeomorphism invariance.
We may thus expect conformal invariance for the action \rf{inva} 
at all distances like for the usual Polyakov string with $\eps=0$ but $\mu_0^2\neq0$.

Some comments about the last statement are in order. The value of $\eps$ as it comes from the Seeley expansion for the Polyakov string is $\propto a^2/\brho$, \ie very small.
This smallness of the coupling is compensated in the diagrams by the largeness of the
integrals, so an uncertainty of the type $\eps\times 1/\eps$ emerges~\cite{Mak21} that 
gives a finite result like in anomalies. Thanks to diffeomorphism invariance one can change 
the value of $\eps$ by adding a constant to $\vp$, so superficially its precise value 
makes not much sense.
However, if we have external dimensionful parameters like momenta $p_a$, the invariant
value $\eps p^2$  does not change and remains small for $p^2\ll\eps^{-1}$.
Thus the above statement about conformal invariance means that it is promoted also to 
very small distances $\sim \sqrt{\eps}$ or very large momenta $p^2\sim 1/\eps$.
This fits the Lilliputian scaling limit~\cite{AM16} which is required to have a stringy
continuum limit of the regularized string. Thereby $\eps p^2$ can still be adjusted as
a small parameter in spite of  virtual momenta $p^2\sim 1/\eps$, so 
in general we cannot drop the terms with $\eps$ in the formulas below.

\section{Conformal coordinates \boldmath{$z$} and \boldmath{$\bar z$}\label{s:III}}

A great simplification of the above formulas occurs when using the conformal coordinates $z$ and $\bz$, where the flat metric tensor reads
\be
\hat g_{ab}=\left(
\begin{array}{cc}
0 ~&~\frac12 \\
\frac12 ~&~ 0\\
\end{array}
\right),\qquad
\hat g^{ab}=\left(
\begin{array}{cc}
0 ~&~2 \\
2 ~&~ 0\\
\end{array}
\right),
\ee
resulting in $\hat R=0$.
The action \rf{inva} then becomes local in the gauge \rf{confog}
\be
{\cal S} =\frac 1{4\pi b_0^2}\int \Big[  \partial \vp  \bp\vp +\frac{\mu_0^2 }2\e^{\vp}
+ 4\eps \e^{-\vp}(\p\bp\vp)^2
-4G \eps \e^{-\vp}\partial \vp\bp\vp \p\bp\vp
\Big].
\label{NG1}
\ee
The first two terms on the right-hand side are nothing but 
 the Liouville action~\cite{Pol81} which is associated with the
conformal anomaly. The field $\vp$ was thus named the Liouville field.
Higher terms are suppressed for
smooth metrics as powers of $\eps \Delta \vp$.

The role of the parameter $\eps= a^{2}/\brho$ 
is two-fold. On the one hand it provides an ultraviolet worldsheet cutoff
as usual for scalar theories with higher derivatives in the action. On the other hand
it has the meaning of a coupling constant for self-interaction of $\vp$ generated by 
the action \rf{NG1} which is specific for theories with diffeomorphism invariance.
As shown in \cite{Mak21}, these two roles played by $\eps$ result in 
a new kind of string anomalies which emerge after doing uncertainties like 
$\eps \times \eps^{-1}$.
Thus, typical $\vp$'s which are essential in the
path integral  are not smooth and have $-\Delta \vp\sim \eps^{-1}$, 
so the higher-derivative terms revive. 
The Liouville action is thus only
an effective action governing smooth classical configurations.
It was also argued that
the next-order in $\eps$ terms omitted in \rf{NG1},  which depend in general on 
the regularization applied,
do not change the results because we are dealing with anomalies.


The corresponding $T_{zz}$ and $T_{z\bar z}$ components of 
the energy-momentum tensor \rf{Tab}  read%
\footnote{For $G=0$ the energy-momentum tensor \rf{Tzz} coincides modulo normalization
with the one given by Eq.~(21) of
\cite{KN93} if we substitute  there the auxiliary field $\chi$  by $\i\eps R/b_0^2$.
That paper deals with an opposite limit of $\eps\to\infty$ in our notations, where the first
term on the right-hand side of \eq{inva} can be omitted.}
\begin{subequations}
\bea
-4 b_0^2 T_{zz}&=&(\p \vp)^2 -2\eps \p \vp \p \Delta \vp -2\q\p^2 (\vp - \eps\Delta \vp)
-G \eps (\p \vp)^2 \Delta \vp+ 4G\eps\p\vp \p(\e^{-\vp}\p \vp \bp \vp)  \non
 && 
- 4G\q\eps \p^2(\e^{-\vp}\p \vp \bp \vp)
 +  G\q \eps\partial(\p \vp \Delta\vp)+
  G \q\eps \frac 1{\bp}\p ^2 (\bp \vp \Delta\vp),
  \label{Tzz}\\
  T_{z\bz}&=&0,
  \label{Tzbz}
\eea
\end{subequations}
where we used the notation $\partial\equiv \partial/\p z$ and
$\bar \partial\equiv \partial/\p\bz$  and $\Delta=4\e^{-\vp}\p  \bp $.
\rf{Tzz} obeys the conservation law
\be
\bar \partial T_{zz}=0.
\label{conserv}
\ee
Both \eq{Tzbz} and \eq{conserv} hold
owing to the classical equation of motion \rf{cemG}.  
In the quantum case it is replaced by the Schwinger-Dyson equation
\be
\hbox{left-hand side of \eq{cemG}}\stackrel{{\rm w.s.}}=8\pi b_0^2 \frac {\delta}{\delta \vp},
\label{SDe}
\ee
where the equality is undersood in the weak sense, \ie under the sign of averages.
Equation~\rf{cemG} is recovered in the classical limit $b_0^2\to0$.

Equation~\rf{Tzz} can be directly obtained from the action~\rf{NG1} by making 
the conformal-type change
\be
\vp \to \vp + \partial \xi + \xi \partial \vp
\label{conftype}
\ee
with $\xi$ depending on both $z$ and $\bar z$. The change of the action \rf{NG1} 
is then 
\be
\delta {\cal S} = 4\pi \int T_{zz} \bar \partial \xi 
\label{newT}
\ee
with $T_{zz}$ given by \eq{Tzz} because $T_{z \bz}=0$ for traceless $T_{ab}$.
Notice that the additional ``nonminimal'' terms coming from the variation of $\hat R$
in \rf{inva} with respect to $\hat g^{zz} $ are also reproduced.
 For the conformal transformation $\xi$ depends only on $z$,
so the variation \rf{newT} identically vanishes, proving the invariance of the action.

\section{DDK for the Polyakov string with curvature squared\label{s:G0}}

Let us begin with repeating the DDK construction for $G=0$, \ie for the Polyakov string with
the curvature squared added to the Liouville action. The first simple observation is that the free
propagator is $8\pi b_0^2  G_\eps(z)$ where
\be
G_\eps(z)=-\frac1{2\pi} \left[\log \big( \sqrt{z\bz}\,{\cal R} \big)+
K_0 \left(\sqrt{\frac {z\bz}\eps}\right)
\right]
\label{Geps}
\ee
with ${\cal R}$ being an infrared cutoff, which obeys
\be
(-4 \p\bp +16 \eps \p^2 \bp^2) G_\eps(z) =\delta^{(2)}(z)
\ee
and we have 
\be
(\p -4 \eps \p^2 \bp) G_\eps (z)=-\frac1{4\pi z}
\ee
for any $\eps$. For this reason the conformal Ward identities are the same as the usual ones
for $\eps=0$.

For convenience of the one-loop 
computation we split $T_{zz}$ given by \eq{Tzz} 
into three pieces
\be
T_{zz}=T_{zz}^{(1)}+T_{zz}^{(2)}+T_{zz}^{(3)} +{\cal O}(\vp^4),
\label{split}
\ee
where
\begin{subequations}
\bea
T_{zz}^{(1)}&=&\frac{1\q}{2 b_0^2} \Big[\p^2 \vp -4\eps \p^3 \bp \vp \Big],\label{T1}\\
T_{zz}^{(2)} &=&\!-\frac1{4 b_0^2} \Big[
(\p \vp)^2 -8\eps \p \vp \p^2 \bp \vp -8 \q\eps \p^2 (\vp\p\bp \vp)
-4G\q \eps \p^2(\p \vp  \bp \vp) + 2 G\q\eps \p\bp (\p \vp)^2 \nonumber 
  \\*[-2mm] && \hspace*{11mm}\hspace*{.2cm}1)\hspace*{1.2cm}2)\hspace*{2.1cm}3)
  \hspace*{1.9cm}4)\hspace*{2.1cm}5)
 \non&& \hspace*{11mm} +2
  G\q \eps \frac 1{\bp}\p ^3 (\bp\vp) ^2\Big]  \label{T2} 
  \\*[-2mm] && \hspace*{11mm} \hspace*{14mm} 6) \nonumber
  \\ 
T_{zz}^{(3)}&=& \!-\frac1{4 b_0^2} \Big[8\eps \p \vp \p (\vp \p \bp \vp) +4\eps \p^2 (\vp^2 \p\bp \vp)-4G \eps (\p \vp)^2 \p\bp \vp 
+4G\eps\p\vp \p(\p \vp \bp \vp)  \nonumber  \\*[-2mm]  &&
\hspace*{15mm}\hspace*{.4cm}7)\hspace*{2.3cm}8)\hspace*{2.2cm}9)
  \hspace*{2.1cm}10)\non
 && \hspace*{11mm}+4G\q\eps \p^2({\vp}\p \vp \bp \vp)
 -4  G \q\eps\partial(\p \vp \vp \p\bp\vp)-4
  G \q\eps \frac 1{\bp}\p ^2 (\bp \vp \vp \p\bp\vp)\Big] .  \label{T3}
  \\*[-2mm]  
  &&  \hspace*{11mm}\hspace*{1.2cm}11)\hspace*{2.3cm}12)\hspace*{2.7cm}13)
 \nonumber
\eea
\label{T123}
\end{subequations}
\!Quartic and higher terms in \eq{split} will be not essential with the one-loop accuracy.
We set $G=0$ in these formulas for the calculation below in this section.

Analogously, we expand the action \rf{NG1} in $\vp$ to explicitly write down the involved
cubic and quartic self-interaction of $\vp$:
\be
{\cal S}={\cal S}^{(2)}+{\cal S}^{(3)}+{\cal S}^{(4)}+{\cal O}(\vp^5)
\ee
with
\begin{subequations}
\bea
{\cal S}^{(2)}&=&\frac 1{4\pi b_0^2} \int \Big[ \p\vp \bp \vp +4 \eps (\p\bp \vp)^2 \Big],
 \label{S2}\\
{\cal S}^{(3)}&=&-\frac 1{\pi b_0^2} \int \Big[ \eps \vp (\p\bp \vp)^2 
+G\eps \p \vp \bp \vp \p\bp \vp\Big], \label{S3}\\
{\cal S}^{(4)}&=&\frac 1{\pi b_0^2} \int \Big[ \frac 12 \eps \vp^2 (\p\bp \vp)^2 
+G\eps \vp\p \vp \bp \vp \p\bp \vp\Big]
\label{S4}
\eea
\label{S123}
\end{subequations}
for $\mu_0=0$.

\subsection{Conformal weight at one loop }

\begin{figure}
\includegraphics[width=10cm]{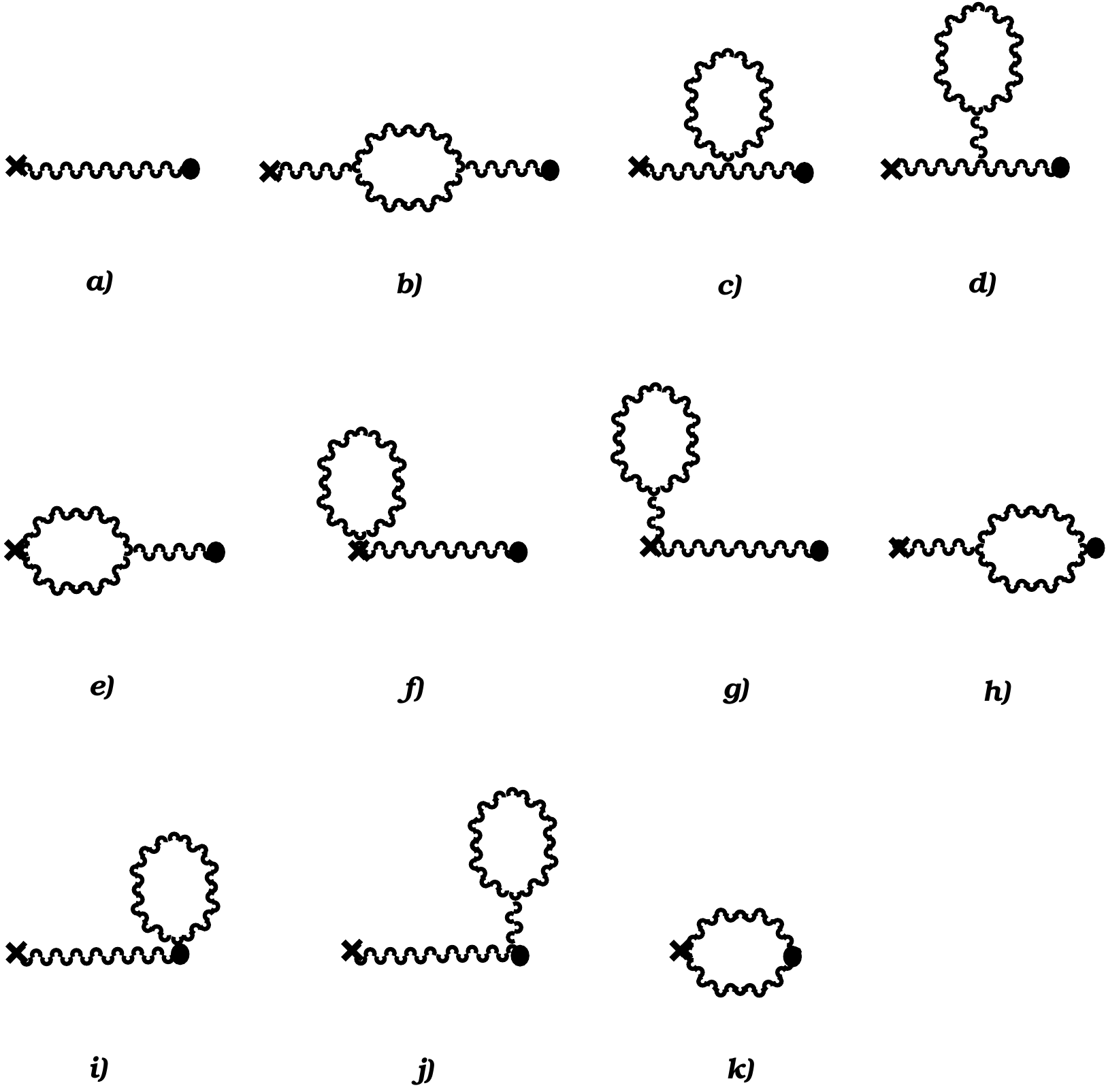} 
\caption{One-loop diagrams contributing to  the operator-product expansion of
$T_{zz}(z) \e^{\vp(0)}$. The points $z$ and 0 are depicted 
by the cross and the dot, respectively.}
\label{fi:one-loop}
\end{figure}

Given \rf{T123} we can apply the conformal field theory technique. 
Let us begin with the calculation of the operator
product $T_{zz}(z)\e^{\vp(0)}$ to find
the conformal weight of the primary operator $\e^\vp$. 
The corresponding  tree and one-loop diagrams are depicted in 
Fig.~\ref{fi:one-loop}, where the wavy lines represent $\vp$.
The diagrams in Figs.~\ref{fi:one-loop}a), b), c), d), h), i), j) come from $T_{zz}^{(1)}$ 
given by \eq{T1},
those  in Figs.~\ref{fi:one-loop}e), g), k) come from $T_{zz}^{(2)}$ given by \eq{T2} and
that in Fig.~\ref{fi:one-loop}f) comes from $T_{zz}^{(3)}$ given by \eq{T3}.
The diagrams in Figs.~\ref{fi:one-loop}b), c), d) are just the one-loop renormalization of
the propagator in Fig.~\ref{fi:one-loop}a), changing $b_0^2\to b^2$
in the effective action
\be
{\cal S}^{({\rm eff})} =\frac 1{4\pi b^2}\int  \p \vp \bp \vp.
\label{S0eff}
\ee
The diagrams  in Figs.~\ref{fi:one-loop}h), i), j) describe the 
renormalization of the operator $\e^\vp$, changing $\vp\to \alpha\vp$ in the exponent.
Both renormalizations were already computed at one loop in \cite{Mak21}. 
The diagrams in Figs.~\ref{fi:one-loop}e), f), g)
renormalize $1/b_0^2\to q/b^2$ in $T_{zz}^{(1)}$, so we arrive at
\be
T_{zz}^{({\rm eff})}=-\frac1{4 b^2} \left[ (\p \vp)^2 -2 q \p^2 \vp\right]
\label{Tzzeff}
\ee
for smooth $\vp$.
Therefore, the contribution of the diagrams a) to j) to the conformal weight of 
$\e^{\alpha \vp}$  is described by the quadratic effective 
action \rf{S0eff} and energy-momentum tensor \rf{Tzzeff} and equals $q \alpha$:
\be
\frac 12 8\pi q\alpha \p^2 G_\eps(z) \stackrel{\eps\to0}\to\frac{q\alpha}{z^2}.
\label{newDDK}
\ee


 The last diagram Fig.~\ref{fi:one-loop}k) gives the usual contribution $-b^2\alpha^2$ which
 comes from the term 1) in \rf{T2}: 
\be
1)=-\frac14 (8\pi)^2 b^2\alpha^2\Big( \p G_\eps(z) \Big)^2 \stackrel{\eps\to0}\to-\frac {b^2\alpha^2}{z^2}.
\label{newDDK2}
\ee
The two other terms 2) and 3) 
 (remember $G=0$ in this section)
 are proportional to $\eps$ which
 cannot be compensated by the derivatives because the second term in \rf{Geps} is
 exponentially small as $z\bz\gg\eps$.
 Thus the sum of all diagrams in Fig.~\ref{fi:one-loop} is just the same as in the DDK case and
 given by the sum of \rf{newDDK} and \rf{newDDK2}. This yields the equation
\be
q\alpha - b^2 \alpha ^2=1
\label{DDK2}
\ee
which states that the conformal weight of $\e^{\alpha \vp}$ is 1.

I would like to emphasize that the renormalization of $b^2$, $q$ and $\alpha$
was not introduced ad hoc but rather derived from the presence of the self-interaction of
the Liouville field $\vp$ owing to the presence of the curvature squared in the action.
I shall check below these results by explicitly computing $b^2$, $q$ and $\alpha$
at one loop directly from the action without using the methods of conformal field theory.

\subsection{Central charge at one loop}

An analogous analysis can be given for the operator product $T_{zz}(z) T_{zz}(0)$
which determines the central charge. The corresponding one-loop diagrams are 
the same as depicted in Fig.~\ref{fi:one-loop}. The classification of the diagrams is also similar.
The diagrams b), c) and d) are the one-loop correction to the tree propagator  a) 
and renormalize $b_0^2\to b^2$. The diagrams e), f) and g) contribute to $q/b^2$ in 
$T_{zz}^{\rm (eff)}(z)$ given by \eq{Tzzeff}. Analogously, the diagrams h), i), j) contribute
to $q/b^2$ in $T_{zz}^{\rm (eff)}(0)$.
All together the diagrams a) to j) give
\be
{\rm tree}=\frac{q^2}{4b^4} \LA \p^2\vp(z) \p^2 \vp(0) \RA=\frac{q^2}{b^2} 2\pi 
\p^4 G_\eps(z) \stackrel{\eps\to0}\to3 \frac{q^2}{b^2} \frac1{z^4}
\ee
which contributes $6q^2/b^2$ to the central charge.

The remaining computation of the diagram k), representing
 $T_{zz}^{(2)}(z) T_{zz}^{(2)}(0)$, is again the same as usual. Only the first term in \rf{T2}
  \be
\hbox{1)-1)}=\frac{1}{16b^4} \LA \p\vp(z)\p\vp(z)  \p\vp(0)\p\vp(0) \RA =\frac 18 (8\pi)^2
 \Big(\p^2G_\eps(z) \Big)^2 \stackrel{\eps\to0}\to \frac1{2z^4}
 \label{38}
\ee
is essential because $\eps$ cannot be compensated by the derivatives for $z\bz\gg \eps$
for the other terms,
 so the contribution of the diagram k) 
to the central charge is 1. Summing all the diagrams
 a) to k) we finally obtain for the total central charge
\be
-\frac6{b_0^2}+1+ \frac{6q^2}{b^2} =0,
\label{DDK1}
\ee
where the first term on the left-hand side is the contribution from $X^\mu$ and ghosts.
The DDK equation \rf{DDK1} represents the vanishing of the total central charge.

It is worth noting that both \eq{DDK1} and \eq{DDK2} are derived at one loop, \ie to
the order ${\cal O}(b_0^2)$ for \eq{DDK1} and  ${\cal O}(b_0^4)$ for \eq{DDK2}.
I shall give below some arguments why these equations can be exact.

\section{An extension of DDK to the Nambu-Goto string\label{s:V}}

I consider in this section an extension of the results of Sect.~\ref{s:G0} to $G\neq0$.
The appropriate expansion of $T_{zz}$ in $\vp$ is shown in \eq{T123}.
The one-loop calculation of the conformal weight of 
the operator $\e^{\vp}$ and the central charge is pretty much similar to that of Sect.~\ref{s:G0}
for $G=0$. The operator products  $T_{zz}(z)\e^{\vp(0)}$ and  $T_{zz}(z) T_{zz}(0)$ 
are given at one loop by the diagrams in Fig.~\ref{fi:one-loop}, accounting for
the additional terms in \rf{T123} and \rf{S123} for $G\neq0$.
In both cases the diagrams a) to j) repeat the consideration of Sect.~\ref{s:G0} and
give the same results $q\alpha$ and $6q^2/b^2$ for the conformal weight and
the central charge, respectively. What is actually to be computed is the contribution
from the diagram k).

\subsection{Conformal weight} 

The analysis of the contribution of the diagram k) to the 
one-loop conformal weight is analogous
to that in Sect.~\ref{s:G0}. The term 1) in \rf{T2} contributes $-b^2\alpha^2$, while for
the terms 2), 3), 4) and 5) the derivatives of $G_\eps(z)$ cannot compensate the factor $\eps$, 
as is immediately seen from \eq{1120}. In momentum space the term 1) contributes $\propto p_z^2/p^2$, where the denominator comes from the integral over
the domain $k^2\sim p^2\ll \eps^{-1}$. Contributions from the terms 2) to 5) is
 $\sim \eps p_z^2$ for $\eps\to0$ and suppressed as  $\eps p^2$.
A special consideration is needed for the term 6) which is
nonlocal. We have in coordinate space
\be
6)=(8\pi)^2 2 G\eps b^2\alpha^2 \p^4 \int G_0(z-\om) \Big(\bp G_\eps(\om)\Big)^2
\stackrel{\eps\to0}\to 2\pi G\eps b^2 \alpha^2 \p^2 \delta^{(2)}(z),
\label{dim6}
\ee
where we used \eq{112d} from the list of formulas in Appendix~A.
It is clear from \eq{dim6} that it is $\sim \eps p_z^2$ in momentum space and 
gives no contribution to the conformal weight which
is therefore given to this order also by \eq{DDK2} like for $G=0$.

\subsection{Central charge}

An analogous computation can be given for the central charge extracted from
the operator product $T_{zz}(z) T_{zz}(0)$. The diagrams of the type in 
Fig.~\ref{fi:one-loop} except for k) gives as $\eps \to 0$ the usual result
\be
2\pi \frac {q^2}{b^2}  \frac{p_z^4}{p^2}  \Longrightarrow  \delta c_q= \frac{6q^2}{b^2}
\ee
as is described by the effective action \rf{S0eff} and the energy-momentum tensor \rf{Tzzeff}.

Again a special treatment is required for the diagram k) which in the usual case, when only
the first term on the right-hand side of \eq{T2} is present
and the 1)-1) contribution  to the central charge is 1:
\be
8\pi^2 \int \frac{\d^2 k}{(2\pi)^2} 
\frac {k_z^2 (p-k)_z^2}{(k^2+\eps k^4)[(p-k)^2+\eps (p-k)^4]}=
\frac\pi 3\frac{p_z^4}{p^2} \Longrightarrow  \delta c =1.
\ee
Only the domain of $k^2\sim p^2 \ll \eps^{-1}$ produces $1/p^2$.

The terms 2) to 5) are $\propto \eps$ that cannot be compensated by the contribution
from the domain $k^2 \sim p^2$. Alternatively, the contribution
from the domain $k^2 \sim \eps^{-1}$ does not produce $1/p^2$.
It is different for the term 6) which includes $1/p^2$ by itself.
Now additional contributions to the central charge come from the mixed 
1)-6) and 2)-6) terms of the diagram in  Fig.~\ref{fi:one-loop}k)
\begin{subequations}
\bea
\hbox{1)-6)}+\hbox{6)-1)}&=&-\frac1{16} 2\cdot2 (8\pi)^2 2G \q\eps \frac {p_z^4}{4p^2}
\int \frac{\d^2 k}{(2\pi)^2} \frac {k^4}{(k^2+\eps k^4)^2 }=
-2 \pi G \q\frac{p_z^4}{p^2} ,~~~
\label{dca}\\
\hbox{2)-6)}+\hbox{6)-2)}&=&-\frac1{16} (8\pi)^2 2G \q\eps \frac {p_z^4}{p^2}
\int \frac{\d^2 k}{(2\pi)^2} \frac {\eps k^6}{(k^2+\eps k^4)^2 }
\non &=&4 \pi G \q\frac{p_z^4}{p^2}
-4 \pi G \q\frac{p_z^4}{p^2} \int \d k^2  \frac {\eps}{(1+\eps k^2) }.
\label{dcb}
\eea
\end{subequations}
The last term is logarithmically divergent what is most probably related to
the subtleties of conformal symmetry for the action \rf{NG1} with $G\neq 0$.
The divergences should cancel each other as is demonstrated
in the next section. 
Ignoring the last term
results in $\delta c= 6 G \q $. Finally, the 6)-6) term is $\sim G^2 \eps^3 p_z^4( p^2)^2$
and does not contribute to $c$.

We can repeat the computation in the coordinate space:
\bea
\hbox{1)-6)+6)-1)}&=&2\cdot \frac 1{16} 
\LA 2 G\eps \partial^3\frac 1{\bp} \bp \vp(z) \bp \vp(z) \p\vp(0)\p\vp(0)
\RA \non &=& (8\pi)^2\frac {G \eps}2  \partial^3 \frac 1{\bp}  \Big(\p\bp G_{\eps}(z) \Big)^2  \stackrel{{\rm\rf{112b}}}\to 
\frac {G \pi}2  \partial^3 \frac 1{\bp}  \delta^{(2)}(z)=-3G \frac1{z^4},
\label{mixed}
\eea
\bea
\hbox{2)-6)+6)-2)}&=&2\cdot \frac 1{16} 
\LA 2 G\eps \partial^3\frac 1{\bp} \bp \vp(z) \bp \vp(z)  (-8 \eps)\p\vp(0)\p^2\bp\vp(0)
\RA \non&=&- (8\pi)^2 4G \eps^2  \partial^3 \frac 1{\bp}  \Big(\p\bp G_{\eps}(z) 
\p^2\bp^2 G_{\eps}(z)\Big) \non
&\stackrel{{\rm\rf{112c}}}\to &
-{G \pi}  \partial^3 \frac 1{\bp}  \delta^{(2)}(z)=6G \frac1{z^4},
\label{mixed62}
\eea
where we have ignored the second term in \rf{112c} causing the logarithmic divergence 
in \rf{dcb}. This is just as Mathematica will do the computation.
A similar analysis shows that
\be
\hbox{3)-6)}\sim G\eps^2 \p^5\bp  \delta^{(2)}(z),\quad
\hbox{4)-6)}\sim G^2 \eps^2 \p^5 \bp  \delta^{(2)}(z),\quad
\hbox{5)-6)}\sim G^2 \eps \p^4 \delta^{(2)}(z)
\ee
as $\eps\to0$.

We thus find the following modification of the DDK equation \rf{DDK1}
\be
-\frac6{b_0^2} +1 +\frac{6q^2}{b^2} +6 G q=0 ,
\label{newDDK1}
\ee
where we multiplied $G$ by $q$ to have an invariant product.
Such a multiplication effects in \rf{newDDK1} the order ${\cal O}(b_0^2)$ or higher 
which is beyond our consideration.
The additional contribution to the central charge, given by  the last term on the left-hand side 
of \eq{DDK2}, is positive for positive $G$, but its negative sign 
for $G<0$ would not spoil positivity of the central charge
because $b_0^2$ is considered small in the perturbative expansion.

It is worth noting once again  
that the introduction of $q$ is due to historical reasons~\cite{Dav88}.
Neither $q$ nor $\alpha$ have much sense separately because they change with changing
the normalization of $\vp$. It is also the case for $G$.
Only the combinations $b\alpha$, $q/b$ and $G q$ make sense. I kept
both $q$ and $\alpha$ for generality. 
We can set for simplicity $q=1$ like in \cite{KN93}.

It is tempting to assume that the one-loop equation \rf{newDDK1} is exact. I shall give below a few
arguments in favor of that but it is merely a speculation.
Setting $q=1$ as already mentioned, we find the solution to Eqs.~\rf{DDK1},
\rf{DDK2}
\bea
\frac1{b^2}=\frac1{b^2_0}-\frac 16 -G,\quad
b\alpha=\frac1{2b}-\sqrt{\frac1{4b^2}-1}.
\label{48}
\eea
It is just the same as DDK with shifted value of $d$:
\be
d\to \tilde d =d+6G.
\label{49}
\ee
All numbers are now real for $d<1-6G$ or $d>25-6G$.

\subsection{The string susceptibility index}

In the DDK approach the string susceptibility index
 simply follows from a uniform dilatation of space, which means adding a constant
 to $\alpha\vp$. Then the higher-derivative terms in the action \rf{A13} are not
 important because the configuration is smooth, while the first 
  term in the brackets involves the topological Gauss-Bonnet term, explaining why
the Euler characteristic $2-2h$ has appeared in $\gamma_{\rm str}$.
The string susceptibility index is thus given by
\be
\gamma_{\rm str}=(1-h)\frac q{\alpha b^2} +\gamma_1
\label{gstr}
\ee
where $\gamma_1$ depends on the boundary condition. For a closed string in $R_d$ we
have $\gamma_1=2$ yielding \eq{ggg}.

The presence of $G$ in \eq{48} may drastically alter the conclusion about 
the existence of the $d=1$ barrier for strings inferable  from \rf{ggg}.
For $G\neq 0$ we may expect from \rf{gstr} the same formula~\rf{ggg} but with the
shift of $d$ given by~\rf{49}.
For $G<0$ this shifts the barrier to $d>1$.
For certain values of $G$ the domain where $\gamma_{\rm str}$ is real
may include $d=4$, while
for $G=-1/6$ the barrier would range from $2$ to $26$ which might seem natural.
A very special case is $G=2$ when $\gamma_{\rm str}=\gamma_1$ for any $h$, 
like for a torus or a cylinder, independently of $d$ for $d<13$. One again all this numerology
relies on the assumption that \eq{newDDK1} is exact while it is actually derived at one loop.

\section{Salieri's check of DDK\label{s:sal}}

As I already pointed out, the beautiful calculation described above is based on the
conformal field theory technique and deals with finite quantities.
It does not need a regularization. However, direct quantum field theory computations
require regularization of emergent divergences. 
As such I use below the Pauli-Villars regularization to check DDK at one loop.

\subsection{Pauli-Villars regulators as conformal fields}

To implement the Pauli-Villars regularization, we add to \rf{inva} or \rf{A13} the following action for the regulator field $Y$:
\be
{\cal S}^{({\rm reg})} =\frac 1{16\pi b_0^2}\int \sqrt{g}\left[ g^{ab}\partial_a  Y \partial_b Y +M^2 Y^2+
\eps (\Delta Y)^2 +G\eps g^{ab}\partial_a  Y \partial_b Y R  \right] 
\label{S0reg}
\ee
or
\be
{\cal S}^{({\rm reg})} =\frac 1{4\pi b_0^2}\int \left[ \partial  Y \bp Y  
+\frac{M^2}4 \e^\vp Y^2+
4\eps \e^{-\vp} (\p  \bp Y)^2 -4 G\eps \e^{-\vp}\partial  Y\bp Y\p\bp \vp  \right] 
\label{S0regc}
\ee
for the gauge \rf{confog} in the conformal coordinates. 
The field $Y$ has a very large mass $M$ and obeys wrong statistics to produce 
the minus sign for every loop, regularizing devergences coming from the loops of $\vp$.

To be precise, the introduction of one regulator is not enough to regularize all the divergences.
Some logarithmic divergences still remain. As was pointed out in \cite{AM17c}, the correct
procedure is to introduce two regulators of mass squared $M^2$ with wrong statistics, which can be viewed
as anticommuting Grassmann variables, and one regulator 
of mass squared $2M^2$ with normal statistics. Then all diagrams including quadratically divergent
tadpoles will be regularized. However, for the purposes of computing anomalies via doing uncertainties
$\eps \times \eps^{-1}$ just one regulator $Y$ would be enough. The contribution of the two others is canceled being mass independent.

The contribution of the regulator to the energy-momentum tensor reads
\bea
-4 b_0^2 T_{ab}^{({\rm reg})}&=&\p_a Y \p_b Y -\frac 12 g_{ab} \p^c Y \p_c Y
 -\frac{M^2}2 g_{ab} Y^2-\eps \p_a Y \p_b \Delta Y -
\eps \p_a \Delta Y  \p_b Y \non
 &&  +\eps g_{ab} \partial^c Y\p_c \Delta Y
 +\frac \eps2 g_{ab} (\Delta Y)^2
-G \eps \p_a Y\p_b Y \Delta \vp 
 + G\frac\eps2\p_a\vp \p_b(\p^c Y \p_c Y) \non
 && 
+ G\frac\eps2 \p_a (\p^c Y \p_c Y)\p_b\vp- G\frac\eps2
g_{ab} \p^c\vp \p_c (\p^d Y \p_d Y)  \non &&-
G\q\eps (\p_a\p_b-g_{ab} \p^c \p_c) (\p^c Y\p_c Y).
\label{Treg}
 \eea
It is conserved thanks to the classical equations of motion for $\vp$ and $Y$
\bea
&&\hbox{l.h.s.\ side of \eq{cemG}}+\frac{M^2}2 Y^2-\frac\eps2 (\Delta Y)^2
+G\frac\eps 2 \p^a Y\p_a Y \Delta \vp -G\frac\eps 2\Delta (\p^a Y\p_a Y)\non &&
~~~~=0, \label{emphi} \\
&&-\Delta Y+M^2 Y +\eps \Delta^2 Y+G\eps \p_a (\p^a Y \Delta \vp ) =0,
\eea
respectively, and traceless 
 thanks to the classical equation of motion \rf{emphi} for $\vp$.
Thus the Pauli-Villars regulators are classically conformal fields in spite of they are massive.

This situation seems to be different from the usual one in quantum field theory, 
where an anomaly emerges
if the regularization breaks the classical symmetry. We may thus expect that conformal
symmetry of the classical action \rf{NG1} will be maintained  at the quantum level
for the Pauli-Villars regularization owing  
to diffeomorphism invariance. I shall confirm this below in this section by explicit 
computations at one loop.

For the regulator contribution to $T_{zz}$ we find
\bea
-4 b_0^2 T_{zz}^{({\rm reg})}&=&\p Y \p Y
 -2\eps \p Y \p \Delta Y 
-G \eps \p Y\p Y \Delta \vp 
 + 4G\eps\p\vp \p(\e^{-\vp} \p Y \bp Y) \non && -4 G \q\eps \p^2 ( \e^{-\vp} \p Y \bp Y),
\label{Tzzreg}
 \eea
giving
\begin{subequations}
\bea
-4 b_0^2 T_{zz}^{(2)\,({\rm reg})} &=&\p Y \p Y
 -8\eps \p Y \p^2\bp Y -4 G\q \eps \p^2 (\p Y \bp Y ), 
 \label{Tzzreg2}\\
 -4 b_0^2 T_{zz}^{(3)\,({\rm reg})} &=&
 8\eps \p Y \p (\vp \p\bp Y) -4G\eps (\p Y)^2 \p\bp \vp \non &&
 +4G\eps \p \vp \p(\p Y\bp Y) +4G\eps \p^2 (\vp \p Y \bp Y).~~~~~
 \label{Tzzreg3}
\eea
\end{subequations}

Before doing the computation let me mention once again  that each of
$b^2$, $\alpha$, $q$ and $G$ depends
on the regularization applied. Only the combinations like $b\alpha$, $q/b$ or $Gq$
are universal. These are the ones which do not change under multiplying $\vp $ and 
simultaneously $Y$ by a constant and are determined by the conformal field theory technique.

\subsection{Minimal Polyakov's string}

Let us begin with the simplest case of $\eps=0$ in Eqs.~\rf{NG1}, \rf{Tab} and \rf{S0regc}, \rf{Treg}, \ie the usual Polyakov string (which I call minimal) regularized by Pauli-Villars.
The quadratic part of the one-loop effective action was computed for this regularization 
in Ref.~\cite{AM17c}. For completeness I now repeat this calculation.

\begin{figure}
\includegraphics[width=11cm]{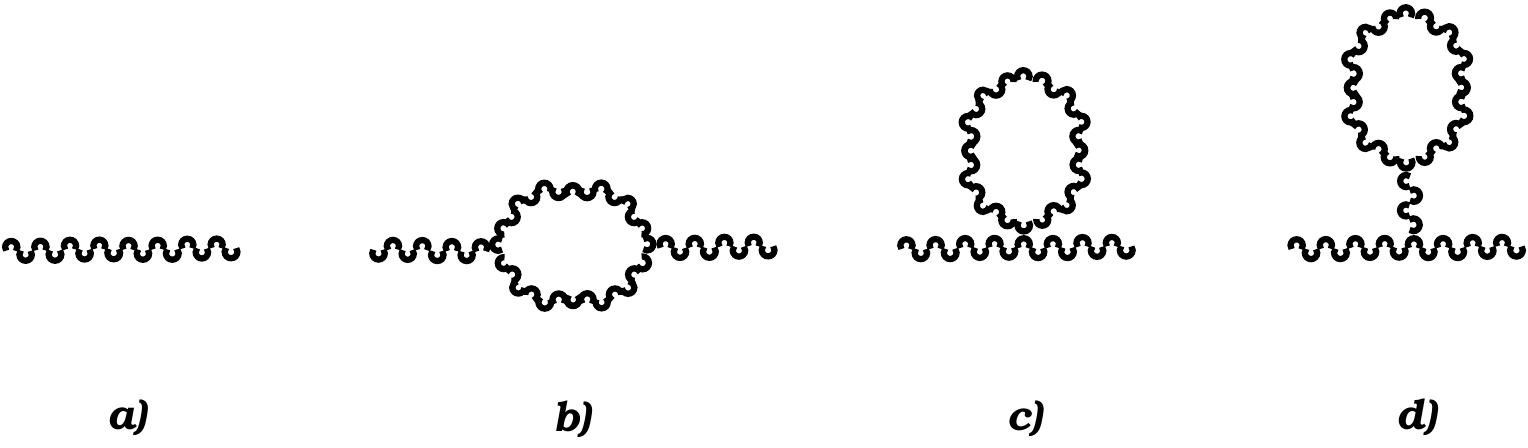} 
\caption{One-loop diagrams for the propagator $\LA{\vp}(z) \vp(0)\RA$.}
\label{fi:gen_pro}
\end{figure}

The one-loop renormalization of the propagator $\LA\vp(-p) \vp(p)\RA$   
 is given by the diagrams in Fig.~\ref{fi:gen_pro}, where the wavy lines represent $\vp$
 and/or the regulators.
 The contribution of the diagram in Fig.~\ref{fi:gen_pro}b)
 to the effective action  reads in momentum space
\bea
\hbox{Fig.~\ref{fi:gen_pro}b)}&=&
\frac14 \!\int \frac{\d^2 k}{(2\pi)^2} 
 \Big\{
\frac{2M^4}{(k^2+M^2)[(k-p)^2+M^2]} \non &&
-\frac{4M^4}{(k^2+2M^2)[(k-p)^2+2M^2]}
\Big\}|\vp(p)|^2 
\stackrel{M\to\infty}\to -\frac{p^2}{96\pi}|\vp(p)|^2,
\label{olb}
\eea
which reproduces the usual conformal anomaly.
The diagrams c) contributes 
\be
\hbox{Fig.~\ref{fi:gen_pro}c)}=-\frac14 \!\int \frac{\d^2 k}{(2\pi)^2} 
 \Big\{
\frac{2M^2}{(k^2+M^2)}
-\frac{2M^2}{(k^2+2M^2)}
\Big\}|\vp(p)|^2=-\frac {M^2}{8\pi} \log 2 \,|\vp(p)|^2 
\ee
which is the renormalization of $\mu^2$ rather than $b^2$. The diagram d) gives no
contribution.
We thus find
\be
\frac{1}{b^2}=\frac1{b_0^2}-\frac16+{\cal O}(b_0^2) .
\label{minb2}
\ee

\begin{figure}
\includegraphics[width=12.5cm]{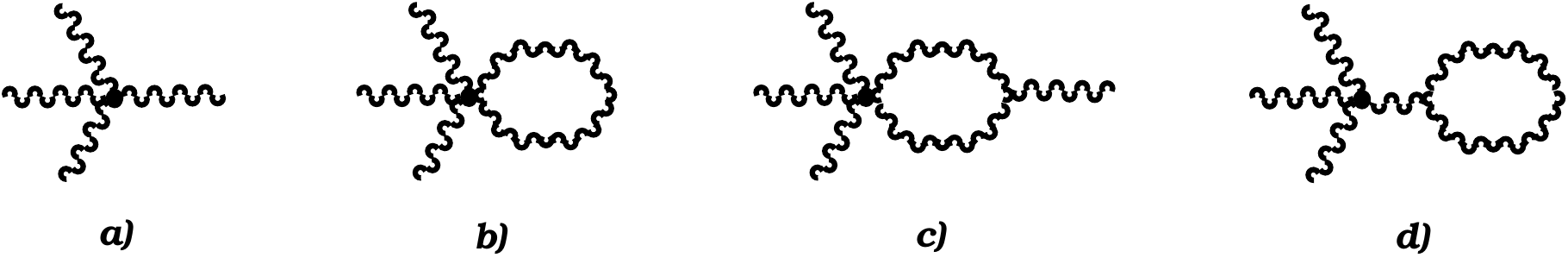} 
\caption{One-loop  diagrams contributing to  the renormalization of  
$ {\e^{\vp(z)}}$. }
\label{fi:a-ren}
\end{figure}

The renormalization of $\alpha$ at one loop is given by the diagrams in Fig.~\ref{fi:a-ren}.
The diagram b) is the standard one. It is logarithmically divergent and thus gives
\be
\hbox{Fig.~\ref{fi:a-ren}b)}= \frac12 \LA \vp(0)^2 \RA =b_0^2\left[\vp(0)+{\rm const.}\right]
\ee
because the coordinate-space propagator logarithmically diverges at coinciding points and
the worldsheet cutoff $\eps\propto a^2\e^{-\vp}$.
We thus find
\be
\alpha=1+b_0^2+{\cal O}(b_0^4).
\label{minal2}
\ee
The diagrams c) and d) are vanishing for $\eps=0$.

\begin{figure}
\vspace*{7mm}
\includegraphics[width=10cm]{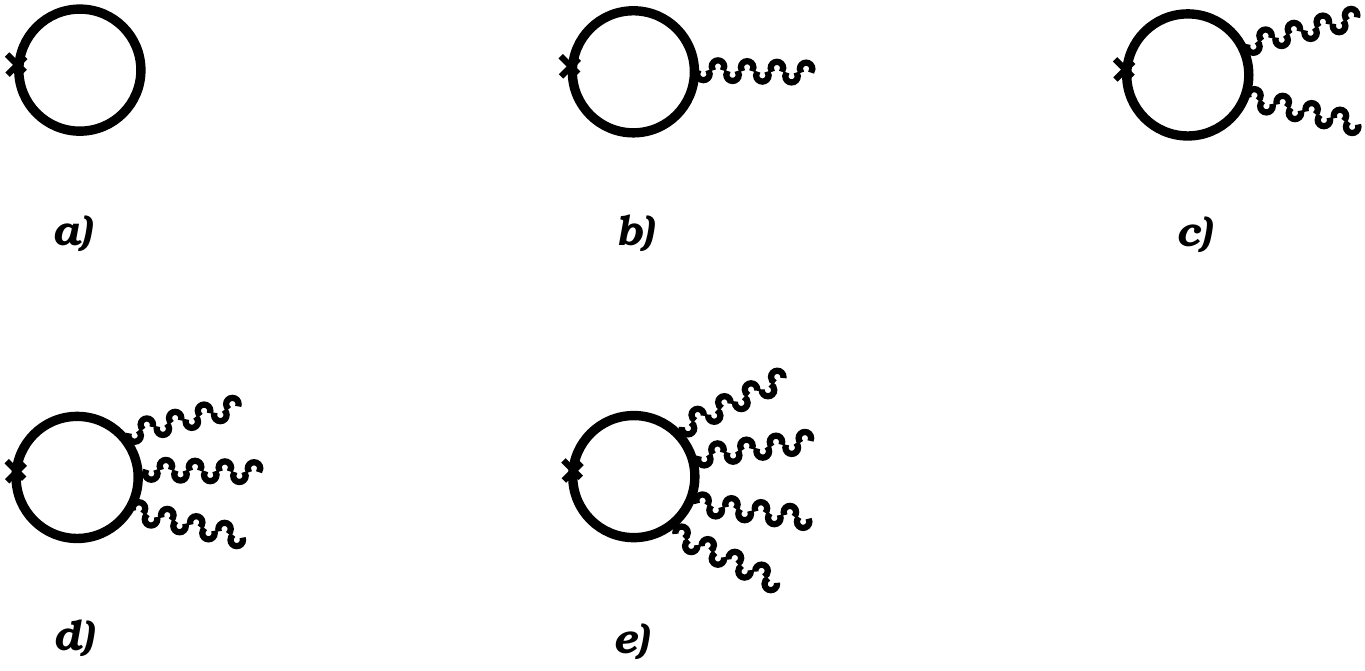} 
\caption{One-loop  diagrams contributing to  the renormalization of  
$ T_{zz}$. }
\label{fi:Tzz}
\end{figure}

The one-loop  renormalization of $T_{zz}$ 
for the minimal Polyakov string, which simplifies  for $\eps=0$ to 
\be
T_{zz}= \frac1{2 b_0^2}\Big[\partial^2 \vp -\frac12(\p\vp)^2 \Big],\qquad
 T_{zz}^{({\rm reg})}=-\frac1{4 b_0^2}(\p Y)^2,
 \label{Tzzm}
\ee
was computed in Appendix A.2 of Ref.~\cite{Mak18}.
The corresponding diagrams 
are depicted in Fig.~\ref{fi:Tzz}, where the solid line
represents $Y$ and the wavy lines represent $\e^\vp\!-\!1$. 
The diagrams come from the averaging $T_{zz}^{({\rm reg})} $ in \eq{Tzzm}
[the first term in \rf{Tzzreg}] over $Y$.

For the diagram in Fig.~\ref{fi:Tzz}b)  we have in momentum space explicitly
\bea
\hbox{Fig.~\ref{fi:Tzz}b)}&=&\frac14 (8\pi)\!\int \frac{\d^2 k}{(2\pi)^2} k_z (p-k)_z
 \Big\{
\frac{2M^2}{(k^2+M^2)[(k-p)^2+M^2]} \non && \hspace*{0.3cm}
-\frac{2M^2}{(k^2+2M^2)[(k-p)^2+2M^2]}
\Big\} \vp(p)
\stackrel{M\to\infty}\to \frac{p_z^2}{12} \vp(p).~~~
\label{min1)}
\eea
The sum of this and the other diagrams reads~\cite{Mak18}
\be
\LA T_{zz}^{({\rm reg})}\RA_Y = -\frac1{12}\Big[\partial^2 \vp -\frac12(\p\vp)^2 \Big]
+{\cal O}(\vp^5).
\label{53}
\ee
The computation was done in the order $\vp^4$ 
and illustrates a tremendous 
cancellation of diagrams which happens thanks to diffeomorphism invariance of
the interaction. 

From \eq{53} we get two results. 
Firstly, we find
\be
\frac{q}{b^2}=\frac 1{b_0^2} -\frac 16 +{\cal O}(b_0^2) .
\label{66}
\ee
Comparing with \eq{minb2}, we thus conclude
\be
q=1+{\cal O}(b_0^4) .
	\label{minq}
\ee
Secondly, adding \rf{53} to the tree-level formula \rf{Tzzm}, we obtain 
for the renormalized $T_{zz}$ the same expression \rf{Tzzm} but with the change 
$b_0^2\to b^2$.

The values  \rf{minb2}, \rf{minal2} and \rf{minq}  
satisfy DDK equations~\rf{DDK1}, \rf{DDK2} to the given order.

The discussed one-loop computation of $T_{zz}$ 
for the minimal Polyakov string  is useful in addressing
a very interesting question: why the quadratic effective action of DDK gives exact
results?
In the language of quantum field theory there is a lot of higher loop diagrams with a
tremendous cancellation between them because the interaction is 
diffeomorphism invariant.
These kind of the cancellation of diagrams can be the reason why
one loop is exact for the minimal Polyakov string.

\subsection{Polyakov's string with curvature squared}

Let us now repeat the analysis of the previous subsection for
the case of $\eps\neq 0$ while $G=0$
in Eqs.~\rf{NG1}, \rf{Tab} and \rf{S0regc} and \rf{Treg}, \ie the Polyakov string with the curvature squared added.
As is already mentioned, the
adding of the $R^2$ term to the action provides a kind of
regularization, but not all diagrams
are regularized this way. Say, the diagrams of the tadpole type remain (quadratically) divergent 
and require another regularization, e.g.\ the Pauli-Villars regularization~\cite{AM17c}.
There is also a bunch of diagrams with logarithmic divergences which cancel each other.
The  divergences could spoil conformal invariance,
so we concentrate on how they cancel.


Let us perform an explicit one-loop computation of  $b^2$  and 
$q$ 
for the action~\rf{NG1} with $G=0$.

\paragraph{Propagator:}
The quadratic part of the effective action is given at one loop by 
the diagrams in Fig.~\ref{fi:gen_pro}.
The part of the diagram in Fig.~\ref{fi:gen_pro}b) gives the conformal anomaly
\bea
\hbox{Fig.~\ref{fi:gen_pro}b)}|_{\rm anom}&=&
-\frac14 \!\int \frac{\d^2 k}{(2\pi)^2} 
\Big\{ \frac{\eps^2 k^2(k-p)^2}{(1+\eps k^2)[1+\eps(k-p)^2]}  \non && \hspace*{.5cm}
-2 \frac{(\eps k^2 (k-p)^2-M^2)^2}{(k^2+M^2+\eps k^4)[(k-p)^2+M^2+\eps(k-p)^4]}
\non && \hspace*{.5cm}
+\frac{(\eps k^2 (k-p)^2-2M^2)^2}{(k^2+2M^2+\eps k^4)[(k-p)^2+2M^2+\eps(k-p)^4]}
\Big\} |\vp(p)|^2
\non
&\to&
\hbox{Fig.~\ref{fi:gen_pro}b)}\Big|_{\rm div}-\frac{p^2}{96\pi} |\vp(p)|^2.
\label{cano}
\eea
Here and below I drop the terms of the next orders 
in $p^2$  which are suppressed as powers of either $\eps p^2$
or $p^2/M^2$, but keeping arbitrary $\eps M^2$.
In contrast  to  the minimal Polyakov string from the previous subsection,
the diagram in Fig.\ref{fi:gen_pro}b) has now an additional contribution 
\be
\hbox{Fig.~\ref{fi:gen_pro}b)}\Big|_{\rm add}=- \int \frac{\d^2 k}{(2\pi)^2} 
\frac{\eps^2 k^2}{(1+\eps k^2)^2} p^2|\vp(p)|^2
\label{57}
\ee
which is logarithmically divergent.

The diagram in Fig.~\ref{fi:gen_pro}c) also has a (regularized) divergent part
\bea
{\rm Fig.~\ref{fi:gen_pro}c}\Big|_{\rm div}&= & \frac14
\int \vp^2
\int \frac{\d^2 k}{(2\pi)^2}\left[ \frac{\eps k^4}{(k^2+\eps k^4)}
-2\frac{\eps k^4+M^2}{(k^2+M^2+\eps k^4)}\right. \non && \hspace*{2.7cm}\left.
+\frac{\eps k^4+2M^2}{(k^2+2M^2+\eps k^4)}
\right]
\label{tad2c}
\eea
and the additional part
\be
\hbox{Fig.~\ref{fi:gen_pro}c)}\Big|_{\rm add}=\int \frac{\d^2 k}{(2\pi)^2} 
\frac{\eps}{(1+\eps k^2)} p^2|\vp(p)|^2
\label{59}
\ee
which is also logarithmically divergent. The logarithmic divergences cancel 
in the sum of \rf{57}  and \rf{59}.

The sum of the (regularized) divergent parts of the diagrams Fig.~\ref{fi:gen_pro}b) and c) reads
\be
{\rm Fig.~\ref{fi:gen_pro}b}\Big|_{\rm div}+
{\rm Fig.~\ref{fi:gen_pro}c}\Big|_{\rm div}
=- \frac 14 \Lambda^2 |\vp(p)|^2
\label{60}
\ee
with
\bea
\Lambda^2&=&\frac{1}{8\pi \eps} \left[
4\sqrt{4 M^2\eps-1} 
\arctan\left(\sqrt{4 M^2\eps-1}\right) \right. \non &&~~~~~~ \left.-2\sqrt{8 M^2\eps-1}
\arctan \left(\sqrt{8 M^2\eps-1} \right)-\log\frac{M^2\eps}2\right]\!.
\label{La2}
\eea
In the limit $\eps \to 0$ 
\be
\Lambda^2 \stackrel{\eps\to0}\to \frac{M^2}{2\pi} \log 2
\ee
reproducing \eq{olb}.
In the opposite limit $M\to\infty$ we get
\be
\Lambda^2
\stackrel{M\to\infty}\to  \frac1{4\eps}\Big[ {(2-\sqrt{2})M\sqrt{\eps}}
-\frac1{2\pi}{\log\frac{M^2\eps}2}\Big].
\label{epsb2}
\ee

A subject of concern is the tadpole diagram in Fig.\ref{fi:gen_pro}d) which for $\eps\neq0$ gives
a nonvanishing contribution
\bea
{\rm Fig.~\ref{fi:gen_pro}d)}&= &-\frac1{16\pi}  A \, p^2 |\vp(p)|^2 ,\non
A&=&(8\pi)
 \eps \!\int \frac{\d^2 k}{(2\pi)^2}\left[ \frac{\eps k^4}{(k^2+\eps k^4)}
-2\frac{\eps k^4-M^2}{(k^2+M^2+\eps k^4)}
+\frac{\eps k^4-2M^2}{(k^2+2M^2+\eps k^4)}
\right]\! . \non &&
\label{tad2cc}
\eea
It is regularization-dependent but cancels, as will be momentarily shown, in the 
ratio $q/b$. 

Summing up  \rf{cano}, \rf{57}, \rf{59} and \rf{tad2cc}, we find
the one-loop renormalization of $b^2$ 
\be
b^2=b_0^2 +b_0^4 \left(\frac 16 -4 +A \right)+{\cal O}(b_0^6).
\label{bbb}
\ee
Alternatively,
the divergent part~\rf{60}  results in the renormalization of $\mu^2$ rather then $b^2$.


\paragraph{Energy-momentum tensor:} 
The one-loop computation of $q/2b^2$ is
associated with the diagrams in Fig.\ref{fi:one-loop}e), f) and g).
For the diagram e) we have nonvanishing contributions from 
 the first three terms in \rf{T2}. The terms 1) and 2) give finite results and 3) leads us
 to a logarithmic divergence.
 We find in momentum space
\bea
1)&=&\frac14 (8\pi)\!\int \frac{\d^2 k}{(2\pi)^2} k_z (p-k)_z
 \Big\{ \frac{\eps}{(1+\eps k^2)[1+\eps(p-k)^2]} \non &&
-2 \frac{\eps k^2 (k-p)^2-M^2}{(k^2+M^2+\eps k^4)[(k-p)^2+M^2+\eps(k-p)^4]}\non
&& 
+\frac{\eps k^2 (k-p)^2-2M^2}{(k^2+2M^2+\eps k^4)[(k-p)^2+2M^2+\eps(k-p)^4]}
\Big\} \vp(p)
\to \frac{p_z^2}{12} \vp(p),
\label{1)}
\eea
\bea
2)&=&\frac14 (8\pi) 2\eps\!\int \frac{\d^2 k}{(2\pi)^2} k_z (p-k)_z k^2
 \Big\{ \frac{\eps}{(1+\eps k^2)[1+\eps(p-k)^2]} \non && \hspace*{1cm}
-2 \frac{\eps k^2 (k-p)^2-M^2}{(k^2+M^2+\eps k^4)[(k-p)^2+M^2+\eps(k-p)^4]}\non
&& \hspace*{1cm}
+\frac{\eps k^2 (k-p)^2-2M^2}{(k^2+2M^2+\eps k^4)[(k-p)^2+2M^2+\eps(k-p)^4]}
\Big\} \vp(p)
\to 0
\label{2)}
\eea
and
\be
3)=\frac 12 (8\pi) \eps^2 \!\int \frac{\d^2 k}{(2\pi)^2 }\frac{k^2}{(1+\eps k^2)^2} p_z^2 \vp(p)=\int \d k^2\frac{\eps^2 k^2}{(1+\eps k^2)^2} p_z^2 \vp(p),
\label{3)}
\ee
where we have again dropped the terms of the next orders  in $p^2$.
In \rf{1)} and \rf{2)} we have also added the regulator contribution  of the first two terms in \rf{Tzzreg2}.


The integral in \rf{3)} is logarithmically divergent what is alarming. There are no
regulator contributions of this kind. But we should not forget about the contribution from
$T_{zz}^{(3)}$ coming from the term
\be
-\frac 14 8\eps \LA \vp \p\bp \vp \p^2 \vp 
\RA
= -\frac 12 (8\pi) \eps \int \frac{\d^2 k}{(2\pi)^2 }\frac{1}{(1+\eps k^2)} p_z^2\vp(p).
\label{4)}
\ee
The sum of \rf{3)} and \rf{4)} is finite.

It is not yet the whole story because of the tadpole diagram in Fig.~\ref{fi:one-loop}g) which
comes from the term 3) in \rf{T2}.
Its contribution reads
\bea
{\rm Fig.~\ref{fi:one-loop}g)}&= &\frac 14 p^2_z \vp(p) A\,
\label{tadT2}
\eea
with $A$ given by \eq{tad2cc}.

Summing up \rf{1)}, \rf{2)}, \rf{3)} and \rf{tadT2}, we obtain
\be
\frac q{b^2}=\frac{1}{b_0^2}-\frac 16 +2 -\frac 12 A
\label{qbb}
\ee
in the formula \rf{Tzzeff} for $T_{zz}^{({\rm eff})}$.
This is to be multiplied by $b$, given at one loop  by \eq{bbb}, to get the
ratio $q/b$ as is prescribed by the little formula
\be
\frac{q}b=\frac q{b^2}\times b .
\label{little}
\ee
The tadpole contributions cancel in the product as is already said
and we arrive again at
\be
\frac {q^2}{b^2} =\frac 1{b_0^2} -\frac{1}{6}  +{\cal O}(b_0^2)
\label{qb}
\ee
which is the same as for the minimal Polyakov string from the previous subsection.

The value \rf{qb} satisfies the DDK equation \rf{DDK1} to the order ${\cal O}(b_0^2)$.
The reason for this is of course that conformal invariance is maintained at one loop and
the contribution of the diagram in Fig.~\ref{fi:one-loop}k) is simply given by \eq{38}
because the contribution of the regulators vanishes.

A miniconclusion of this subsection is that the divergences which might spoil conformal 
invariance indeed cancel at one loop.
I would like to emphasize once again that the renormalization of $b^2$, $q$ and $\alpha$
was not introduced ad hoc but rather derived from the presence of the interaction between
the Liouville field $\vp$ and the regulator $Y$ as prescribed by diffeomorphism invariance.
The ratio $q/b$ which is explicitly computed at one loop coincides with the DDK 
result to this order.
I have made also a few more computations to convince myself that everything does work
but I do not think they are instructive. 


\subsection{The Nambu-Goto string}

Now it is the turn of the case $G\neq0$ which is associated with the Nambu-Goto string.
Additional contributions to $b^2$ coming from the presence of the $G$-term in the action are
\be
{\rm Fig.~\ref{fi:gen_pro}b)}=-2 b_0^4 G \eps \!\int \! \d k^2\! \left[
\frac{\eps k^6}{(k^2+\eps k^4)^2}-\frac{2(\eps k^4-M^2)k^2}{(k^2+M^2+\eps k^4)^2}
+\frac{(\eps k^4-2M^2)k^2}{(k^2+2M^2+\eps k^4)^2}
\right]\!,
\label{b)}
\ee
\be
{\rm Fig.~\ref{fi:gen_pro}c)}=2 b_0^4 G \eps \int \d k^2 \left[
\frac{2k^2}{(k^2+\eps k^4)}-\frac{2k^2}{(k^2+M^2+\eps k^4)}
+\frac{k^2}{(k^2+2M^2+\eps k^4)}
\right]\!,
\label{c)}
\ee
\bea
{\rm Fig.~\ref{fi:gen_pro}d)}&=& -b_0^4 G \eps \int \d k^2 \left[
\frac{\eps k^4}{(k^2+\eps k^4)}-\frac{2(\eps k^4-M^2)}{(k^2+M^2+\eps k^4)}
+\frac{(\eps k^4-2M^2)}{(k^2+2M^2+\eps k^4)}
\right] \non &=&-\frac 12 b_0^4  GA.
\label{d)}
\eea
The sum of \rf{b)} and \rf{c)} is rather simple
\be
{\rm Fig.~\ref{fi:gen_pro}b)+Fig.~\ref{fi:gen_pro}c)}=2 b_0^4 G  \int 
\d k^2\frac{\eps }{1+\eps k^2}.
\label{bc)}
\ee
The tadpole contribution \rf{d)} is  already familiar from \eq{tad2cc}.

Summing \rf{d)}, \rf{bc)} with \rf{bbb} we obtain
\be
b^2=b_0^2 +b_0^4 \left(\frac 16 -4 +A +2 G \int \d k^2 \frac \eps{(1+\eps k^2)}-
\frac 12 G A\right)+{\cal O}(b_0^6).
\label{bbbG}
\ee

In the ratio $q/b$ the contribution of this tadpole is canceled again by the tadpole 
depicted in Fig.~\ref{fi:one-loop}g) which comes from the term 6) in \rf{T2}.
For its contribution to $q/2b^2$ we have
\be
{\rm 6)}= \frac 14 G \eps \int \d k^2 \left[
\frac{\eps k^4}{(k^2+\eps k^4)}-\frac{2(\eps k^4-M^2)}{(k^2+M^2+\eps k^4)}
+\frac{(\eps k^4-2M^2)}{(k^2+2M^2+\eps k^4)}
\right] =\frac 18 GA.
\label{d)6)}
\ee
Multiplying  \rf{d)6)} by 2, we obtain precisely 1/2 of \rf{d)}
with the opposite sign. Adding the two, as is prescribed by \eq{little},
we obtain
\be
2 \rf{d)6)} +\frac1{2b_0^4}\rf{d)}=0
\ee
and see a cancellation of the tadpole diagrams in the ratio $q/b$.

It still remains to compute the diagrams in Figs.~\ref{fi:one-loop}e) and f) and to
 show how the logarithmic divergence in \rf{bc)} is canceled. Let us begin with the diagram 
 in Fig.~\ref{fi:one-loop}f) where nonvanishing contributions come from  
the terms 10), 11), 12), 13)  in \rf{T3}. We  have
\bea
&&10)=12)=13)=-\frac 14 G\eps \int \d k^2 \frac {1}{(1+\eps k^2)}, \\
&&11)= -\frac 14 G\eps \int \d k^2 \left[\frac {2}{(1+\eps k^2)} -\frac {2k^2}{(k^2+M^2+\eps k^4)}
+\frac {k^2}{(k^2+2 M^2+\eps k^4)}
\right]\!,~~
\label{11)T3}
\eea
where we added  the contribution of this type from \rf{Tzzreg3}.

There are several contributions from the diagram in Fig.~\ref{fi:one-loop}e).
The term 4) together with its regulator counterpart gives 
\be
4)= \frac 14 G\eps \int \d k^2 \left[\frac {\eps k^2}{(1+\eps k^2)^2}
 -\frac {2k^2(\eps k^4 -M^2)}{(k^2+M^2+\eps k^4)^2}
+\frac {k^2(\eps k^4 -2M^2)}{(k^2+2 M^2+\eps k^4)^2}
\right].
\label{4)eps}
\ee
The sum of \rf{11)T3} and \rf{4)eps} is again simple and 
summing up the five contributions we obtain
\be
4)+10)+11)+12)+13)=- G \int \d k^2 \frac {\eps }{(1+\eps k^2)}.
\label{88}
\ee

The terms 1) and 2) also give nonvanishing contributions
\bea
1)&=&-\frac 14 (8\pi)^2\frac1{16\pi} 
G\eps \int \frac{\d^2 k}{(2\pi)^2}
\frac{k_z(p-k)_z \left[ k_\bz (k-p)^2+ (p-k)_\bz k^2\right]}
{(k^2+\eps k^4)[(k-p)^2+\eps (k-p)^4]} p_z \vp(p) \non
&\stackrel{\eps\to0}\to &- \frac14 G p_z^2 \vp(p) 
\label{mi1}
\eea
and analogously
\be
2)\stackrel{\eps\to0}\to -\frac 12
G\int\d k^2
\frac{p_z\eps^2 k^2}
{(1+\eps k^2)^2} p_z \vp(p) 
=  -\frac 12
G\int\d k^2
\frac{\eps}
{(1+\eps k^2)} p_z^2 \vp(p) + \frac12 G p_z^2 \vp(p). 
\label{mi2}
\ee
There is no contribution of this kind from the regulators. 

Finally, all the terms of the type $G^2$ vanish.

Summing \rf{mi1} and \rf{mi2} with \rf{88} and adding to \rf{qbb}, we obtain
\be
\frac q{b^2}=\frac{1}{b_0^2}-\frac 16 +2 -\frac 12 A -\frac 12 G 
- G \int \d k^2 \frac {\eps }{(1+\eps k^2)} +\frac 14 G A 
\label{qbbG}
\ee
or, multiplying by $b$, given by \eq{bbbG}, according to \eq{little} and squaring ,
\be
\frac {q^2}{b^2} =\frac{1}{b_0^2} -\frac 16 -G +{\cal O}(b_0^2).
\label{100}
\ee
This precisely confirms the shift of the central charge in \rf{newDDK1} computed in
 Sect.~\ref{s:V} by using the conformal field theory technique.
 It is also instructive to see that $-G$ in \eq{100}
 came entirely from $q/b^2$, \ie from the renormalization
 of $T_{zz}$, while the role of $b^2$ was to cancel the divergences.

\section{Discussion\label{s:conclu}}

The main result of this Paper is the shift~\rf{49} in
the string susceptibility index \rf{ggg} derived 
at one loop for the action \rf{NG1} using the DDK technique. 
The self-consistent one-loop result \rf{DDK2}, \rf{newDDK1} was promoted to be exact 
which gave  the shift~\rf{49} in the string susceptibility index~\rf{ggg}.
I have several arguments in favor of this speculation but no proof of that.

It is especially interesting what would be the value of $G$ emerging for the 
Nambu-Goto string after the path integration over $\lambda^{ab}$.
It is tempting to assume that the factor $6$ in \eq{newDDK1} is linked to the presence
of 3 fields: $\lambda^{zz}$, $\lambda^{\bz\bz}$ and   $\lambda^{z\bz}$.
To perform the computation of $G$, it may turn out to be useful to use 
the generalization~\cite{DOP82} of the Seeley expansion of the heat kernel.
Alternatively, one can try to compute the conformal weight and the central
charge directly from the emergent action for $\vp$ and $\lambda^{ab}$ along
this line.
One more related problem to think about is what effective action can be associated with DDK
for the Nambu-Goto string? For the Polyakov string, both minimal and with the
curvature squared in the limit $\eps\to0$, it is the quadratic effective action \rf{S0eff}.

An urgent task to do is to find out what simplifications occur with the action~\rf{NG1} 
for $G=2$ -- the case mentioned at the very end of Sect.~\ref{s:V}.
It would most probably have nothing to do with the Nambu-Goto string, 
but the action~\rf{NG1} is interesting by itself as an example of conformal field
theory whose central charge is unexpectedly shifted by the presence of the term
with $G$ in the action that preserves conformal invariance apparently for arbitrary $G$.

\subsection*{Acknowledgement}

I am grateful to V.~Kazakov for bringing Ref.~\cite{KSW} to my attention.

This work was supported by the Russian Science Foundation (Grant No.20-12-00195).

\paragraph*{Note added in Manuscript:} The origin of the logarithmic divergence in \eq{dcb}
and correspondingly in the
one-loop central charge is clarified in the recent publication~\cite{Mak22c}.

\appendix
\section{List of useful formulas with regularized propagator\label{appA}} 

The regularized propagator reads
\be
G_\eps (z) =-\frac1{2\pi} \left[ \log \big(\sqrt{{z\bz}}\,{\cal R}\big)+
K_0 \left( \sqrt{\frac{z\bz}\eps } \right) \right] ,
\label{1120}
\ee
where ${\cal R}$ is an infrared cutoff. Differentiating we obtain
\bea
-4\p\bp G_\eps(z) &=&\frac1{2\pi\eps} K_0 \left(\sqrt{\frac{z\bz}{\eps}}\right)
\stackrel{\eps\to0}\to \delta^{(2)}(z), \label{112a}
\eea
\bea
\eps\Big(4\p\bp G_\eps(z) \Big)^2
&=&\frac1{4\pi^2\eps} K_0^2 \left(\sqrt{\frac{z\bz}{\eps}}\right)
\stackrel{\eps\to0}\to\frac1{ 4\pi} \delta^{(2)}(z)  ,
\label{112b}\eea\bea
\eps^2\Big(4\p\bp G_\eps(z) 16\p^2\bp^2 G_\eps(z)\Big)
&=&\frac1{4\pi^2\eps} K_0^2 \left(\sqrt{\frac{z\bz}{\eps}}\right)-\frac 1{2\pi} \delta^{(2)}(z)
K_0 \left(\sqrt{\frac{z\bz}{\eps}}\right) \non
&\stackrel{\eps\to0}\to &
\frac1{4\pi }\delta^{(2)}(z) +\frac 1{2\pi} \delta^{(2)}(z)
\log\sqrt{\frac{z\bz}{\eps}} , 
\label{112c} \eea\bea
\p^2 \Big(\bp G_\eps(z) \Big)^2&\stackrel{\eps\to0}\to&
-\frac1{ 16\pi} \p\bp\delta^{(2)}(z) ,
\label{112d} \\
\eps \Big( 4 \p^2 \bp G_\eps(z) \Big)^2&\stackrel{\eps\to0}\to& \frac1{24\pi} 
\p^2\delta^{(2)}(z) .
\label{112e}
\eea

\end{document}